\documentclass[%
 reprint,
superscriptaddress,
 amsmath,amssymb,
 aps,
 pra
]{revtex4-2}

\usepackage{makecell}
\usepackage{relsize}
\usepackage{graphicx}
\usepackage{dcolumn}
\usepackage{bm}

\usepackage{color}
\usepackage{amssymb}
\usepackage[unicode=true,pdfusetitle,
 bookmarks=true,bookmarksnumbered=true,bookmarksopen=true,
 breaklinks=true,pdfborder={0 0 1},colorlinks=true]
{hyperref}
\hypersetup{linkcolor=blue,anchorcolor=blue,citecolor=blue,urlcolor=blue}


\begin{document}

\preprint{APS/123-QED}

\title{Correlation visibility and generalized Siegert relation for random light beams
}

\author{Yi Cui}
\affiliation{ 
Faculty of Arts and Sciences, Beijing Normal University, Zhuhai 519087, China
}%
\affiliation{%
School of Physics and Astronomy, Applied Optics Beijing Area Major Laboratory, Beijing Normal University, Beijing 100875, China
}%
\author{Wanting Hou}
\affiliation{%
School of Physics and Astronomy, Applied Optics Beijing Area Major Laboratory, Beijing Normal University, Beijing 100875, China
}%
\author{Jun Xiong}
\affiliation{%
School of Physics and Astronomy, Applied Optics Beijing Area Major Laboratory, Beijing Normal University, Beijing 100875, China
}%
\author{Zhiyuan Ye}
\email{yezy@bnu.edu.cn}
\affiliation{ 
Faculty of Arts and Sciences, Beijing Normal University, Zhuhai 519087, China
}

\date{\today}

\begin{abstract}
Phase difference is central to classical coherence theory.  With the advancement of various light-field modulation techniques, artificially generated pseudo-thermal light sources or random light beams can exhibit exotic wavefront correlation properties. However, such spatial wavefront correlations cannot be fully characterized using the phase difference alone. For instance, for a pair of conjugate pseudo-thermal beams, the spatial wavefronts exhibit a significant anti-correlation, meaning that the sum of their wavefronts tends to be constant. In this work, we propose the concept of degree of wavefront correlation $p^{(1)}$, ranging symmetrically from $-1$ to $+1$, for numerically calculating the wavefront correlation properties among various pseudo-thermal light sources, and the sign (positive or negative) can be used to determine the tendency—whether it leans toward wavefront-difference or wavefront-sum correlation. Numerical results demonstrate that the classical Siegert relation does not apply to pseudo-thermal light sources that exhibit wavefront-sum correlation properties. To address this, we propose a generalization valid for all Gaussian pseudo-thermal light.
Experimentally, we introduce the measurable quantities of correlation visibility $\mathcal{V}_g$ and correlation background $\mu_g$, which form a two-dimensional classification framework $\{\mu_g,\mathcal{V}_g\}$ that enables the experimental characterization of diverse Gaussian pseudo-thermal light using a common-path interferometer and intensity correlation measurement. Furthermore, the correlation visibility $\mathcal{V}_g$ can serve as an observable criterion for a zero-mean, non-circularly symmetric, and jointly Gaussian distribution.
\end{abstract}

\maketitle

\section{Introduction} 

In the classic Young's double-slit interference experiment, the fringe visibility $\mathcal{V}$ was the first quantitatively observable measure of coherence, defined as
\begin{eqnarray}
\mathcal{V}=\frac{ I_{ \mathrm{max}} - I_{ \mathrm{min}}  }{ I_{ \mathrm{max}} + I_{ \mathrm{min}} } ,
\label{eq:1}
\end{eqnarray}
where $I_{ \mathrm{max}}$ and $I_{ \mathrm{min}}$ are the maximum and minimum intensities of the fringe pattern, respectively. 
In 1938, Zernike’s definition of the degree of coherence $\gamma$ is identical to the strength of phase correlations \cite{Zernike1938, Tervo2012},
{\small
\begin{eqnarray}
\gamma=\frac{\left|\langle \tilde{E}^*_1\tilde{E}_2\rangle\right|}{\langle A_1A_2\rangle}=\left|\langle\exp\left[\mathrm{i}(\Phi_2-\Phi_1)\right]\rangle\right|=\left|\langle\exp\left(\mathrm{i}\Delta\Phi\right)\rangle\right|,
\label{eq:2}
\end{eqnarray}}
where $\tilde{E}_j\equiv A_j\exp(\mathrm{i}\Phi_j)$ ($j=1,2$) is the optical field of a quasi-monochromatic scalar field with the real amplitude $A_j$ and phase $\Phi_j$, the asterisk denotes the complex conjugate, and $\langle\cdot\rangle$ denotes the ensemble average over time.
For brevity, all explicit spatial coordinates $\mathbf{r}_j$ are omitted, with the understanding that they correspond to the positions indicated by the subscripts of $\tilde{E}_j$.
The theoretical foundation connecting visibility to the underlying statistical properties of the light field is provided by the normalized first-order coherence function \cite{Glauber1963}
\begin{eqnarray}
g^{(1)}\equiv \frac{ \langle \tilde{E}^*_1\tilde{E}_2\rangle }{\sqrt{\langle I_1\rangle\langle I_2\rangle}}=\frac{ \langle \tilde{E}^*_1\tilde{E}_2\rangle }{\sqrt{ \langle  \tilde{E}^*_1\tilde{E}_1 \rangle \langle  \tilde{E}^*_2\tilde{E}_2 \rangle }},
\label{eq:2a}
\end{eqnarray}
which results in $|g^{(1)}|=\mathcal{V}$. 

In 1956, Hanbury Brown and Twiss (HBT)\cite{R6} revealed the photon bunching effect and the optical intensity correlation in thermal light.
The signature of intensity correlations at two points from two optical fields is $\left \langle  I_1I_2\right \rangle \neq \left \langle  I_1\right \rangle \left \langle  I_2\right \rangle$, and the degree of second-order coherence or the normalized intensity correlation function is defined as \cite{Glauber1963}
\begin{eqnarray}
g^{(2)}(I_1,I_2)=\frac{\left \langle  I_1I_2 \right \rangle }{\left \langle  I_1 \right \rangle \left \langle  I_2 \right \rangle},
\label{eq:2b}
\end{eqnarray}
where this paper does not take into account the relative time delay, that is, $\tau=0$ is the default. The degree of second-order coherence (or intensity correlation degree) $g^{(2)}(0)$ can be utilized to distinguish different types of light sources, such as thermal light ($>$1), laser light (=1), and single-photon sources ($<$1).

The Sigert relation\cite{Ferreira2020} gives 
\begin{eqnarray}
g^{(2)}=1+|g^{(1)}|^2, 
\label{eq:2c}
\end{eqnarray}
which connects the first- and second-order coherence properties of Gaussian thermal light. 
{Thermal light is ubiquitous in nature, ranging from blackbody radiation sources to starlight. The coherence time of true thermal light is typically extremely short, on the order of nanoseconds or even picoseconds, significantly increasing the difficulty of detection. 
In practical scientific research and engineering applications, a quasi-monochromatic laser is often passed through a rotating ground glass to simulate the intensity fluctuations of true thermal light \cite{Martienssen1964}. Such a light source is referred to as pseudo-thermal light, and its equivalent coherence time can be quite long, depending on the rotation speed of the ground glass.
With the advancement of various light-field modulation techniques, in modern optics, a laser is often used in conjunction with a spatial light modulator (SLM) to generate tunable pseudo-thermal light. By adjusting the statistical distribution of the random phase pattern loaded onto the SLM, the higher-order statistical properties of the pseudo-thermal light can be manipulated for various optical applications \cite{Maga2016, YangZ2017, Bender2019, PengD2021, YEzy2022, LeeCH2023}.}

{
The theoretical description of pseudo-thermal light consists of two parts: one is the incident laser field (carrier wave), and the other is the spatially random wavefront introduced by various modulation techniques:
}
\begin{eqnarray}
{
\tilde{E}(\mathbf{r},z,t)=E(\mathbf{r},t)\exp\left[\mathrm{i}(2\pi/\lambda)(z-ct)+\mathrm{i}\varphi_t\right],
}
\label{eq:2d}
\end{eqnarray}
{
where $E(\mathbf{r},t)\equiv A\exp\left[\mathrm{i}\phi(\mathbf{r},t)\right]$ is the random wavefront with the spatial position $\mathbf{r}\equiv(x,y)$ and time $t$, and $\phi(\mathbf{r},t)$ represents the spatially random wavefront that fluctuates over time, introduced by various modulation methods, and it determines the spatial coherence of the pseudo-thermal light; $\lambda$ is the wavelength, and $\varphi_t$ is the random phase of the carrier wave determining the coherence time of the incident laser light.
}

From $\mathcal{V}$ to $\gamma$ and from $g^{(1)}$ to $g^{(2)}$, the phase difference $\Delta{\Phi}$ in Eq.~(\ref{eq:2}) plays a crucial role. 
However, {the existing framework cannot fully characterize the spatial correlation properties of various novel pseudo-thermal light sources. For instance,} in the recently proposed holographic thermal light \cite{YeZ2025, HouW2025, YeZ2026}, the pairwise generated conjugate spatially incoherent fields exhibit {opposite wavefront correlation but identical temporal-frequency term, which is to say that their spatial wavefronts are always complementary to each other, or the sum of their wavefronts $\varsigma\phi\equiv\phi_1+\phi_2$ tends to be constant. When directly calculating the first-order coherence function (Eq.~(\ref{eq:2a})) between this pair of conjugate pseudo-thermal light sources, the temporal frequency terms cancel each other out while the wavefront-difference correlation term $\langle E_1E^*_2\rangle$ vanishes, yielding $g^{(1)}=0$. Yet, there exists a strong second-order correlation signal with a theoretical peak value of $g^{(2)}=2$. This result cannot be interpreted by Eq.~(\ref{eq:2c}) any longer.}
In such a case, {inspired by Refs.~\cite{Barnett1990, Chekhova2016, Fuenzalida2024, Erkmen2008, Erkmen2010}, a natural motivation comes to the definition}
\begin{eqnarray}
h^{(1)}\equiv \frac{\langle E_1E_2\rangle}{\sqrt{\langle I_1\rangle\langle I_2\rangle}},
\label{eq:3}
\end{eqnarray}
which {is complementary to} the first-order coherence function $g^{(1)}$ {for the specific scenario of holographic thermal light or any other pseudo-thermal light beams with wavefront-sum correlations}. Also, a special note throughout this paper is that {the spatial wavefront correlations of $E$ rather than $\tilde{E}$ in pseudo-thermal light lie at the core focus of this study}, so the temporal-frequency aspect of the optical field {from the carrier wave} should be omitted.
{
One should distinguish the fundamental difference between $\langle \tilde{E}_1\tilde{E}_2 \rangle$ and $\langle E_1E_2\rangle$. The former involves the random variable $\varphi_t$ according to Eq.~(\ref{eq:2d}), so mathematically $\langle \tilde{E}_1\tilde{E}_2 \rangle=0$, implying the absence of so-called \textbf{phase-sum} correlation. Since the latter does not involve the carrier-wave term, the value of $h^{(1)}$ can be non-zero mathematically. To avoid confusion, two types of spatial correlations are discussed in this paper: one is the \textbf{wavefront-difference} correlation, which is directly related to $g^{(1)}$, and the other is the \textbf{wavefront-sum} correlation, which is directly related to $h^{(1)}$. Therefore, in the subsequent expressions, the temporal-frequency term in Eq.~(\ref{eq:2a}) is naturally omitted as well.
}


Although $g^{(2)}(0)$ can distinguish between different correlated light sources, it cannot differentiate between conventional thermal light and holographic thermal light, as their theoretical values are both 2. 
It is implied that intensity correlation is fundamentally incapable of distinguishing between these two distinctly opposite types of {wavefront} correlations of $\Delta\phi$ or $\varsigma\phi$.
{Furthermore,} given that considerable experimental evidence has indicated the existence of the {wavefront}-sum correlation (i.e., $|h^{(1)}|\neq0$) in classical linear optical systems{, and that pseudo-thermal light can be flexibly engineered, a new question naturally arises: is holographic thermal light completely incompatible with conventional Gaussian pseudo-thermal light? In other words, can a pair of pseudo-thermal light beams exhibit partial wavefront-sum correlation or partial wavefront-difference correlation, or even a mixture of the two? To address this issue,} this work aims to introduce new metrics for quantitative analysis and experimental characterization of the {wavefront} correlation properties between a pair of {Gaussian pseudo-thermal} light fields. 

\section{Theoretical framework}
\subsection{Degree of {wavefront} correlation} 

\begin{figure*}[htbp]%
\centering
\includegraphics[width=\linewidth]{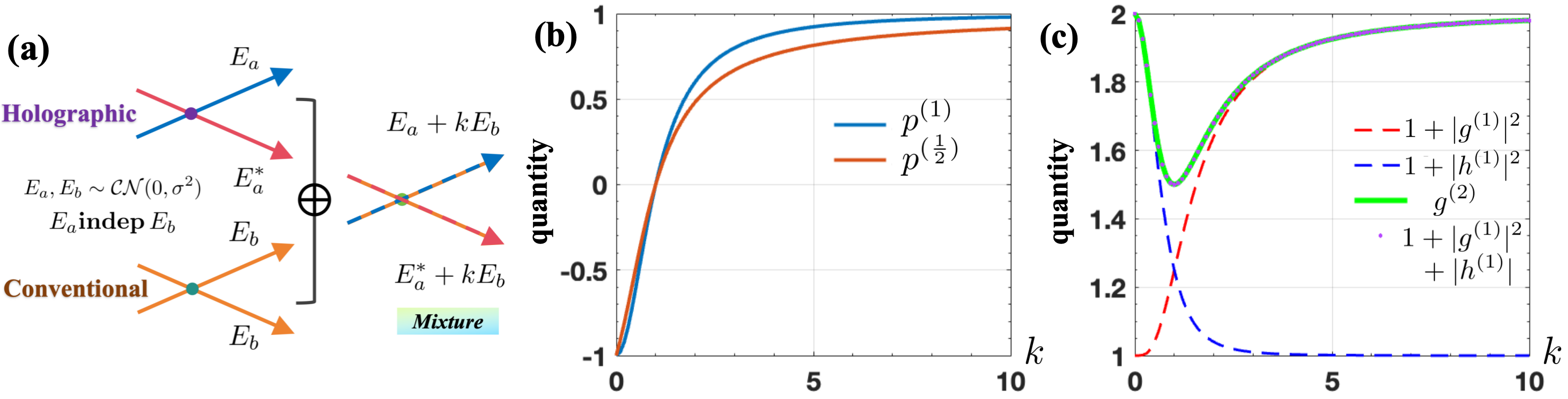}
\caption{
{Schematic diagram (a) of the mixture of holographic thermal light and traditional thermal light.}
Numerical calculation of (b) various correlation degrees for a mixture of holographic and conventional thermal light sources with a scaling factor $k$, and (c) numerical verification of the generalized modified Siegert relation.
{It should be noted that only the wavefront correlations in pseudo-thermal light are considered here, where both $E_a$ and $E_b$ obey complex Gaussian distributions and are statistically independent of each other.}
}\label{fig0}
\end{figure*}

Inspired by Refs.~\cite{Erkmen2008, Erkmen2010}, a generalized form of \textit{degree of {wavefront} correlation} $p^{(1)}$ is theoretically given as
\begin{eqnarray}
p^{(1)}=\left| g^{(1)} \right| - \left| h^{(1)} \right|=\frac{\langle A_1A_2 \rangle}{\sqrt{ \langle A^2_1\rangle \langle A^2_2\rangle }} \left ( \gamma-\eta \right),
\label{eq:4}
\end{eqnarray}
where the real number $p^{(1)}$ is nomarlized symmetrically within the range of $[-1,+1]$, {$\gamma\equiv|\langle\exp(\mathrm{i}\Delta\phi)\rangle|$,} and $\eta\equiv|\langle\exp(\mathrm{i}\varsigma\phi)\rangle|$.
The quantity $p^{(1)}$ is physically defined as the difference between two markedly different degrees of {wavefront} correlation, and the sign of $p^{(1)}$ serves to quantitatively characterize the nature of {wavefront} correlation, namely the compitition of statistical distributions of $\Delta\phi$ and $\varsigma\phi$, between a pair of correlated random light fields at two points.
When $A_1=A_2$, we have $p^{(1)}=\gamma-\eta$.
{It is commonly assumed that $\phi$ follows a uniform distribution from 0 to $2\pi$. So,} $\Delta\phi$ ($\varsigma\phi$) has widths of $4\pi$, and phases differing by $2\pi$ are physically indistinguishable\cite{Barnett1990}. 
Hence, a natural motivation comes to the form of $p^{(\frac{1}{2})}$: 
\begin{eqnarray}
p^{( \frac{1}{2} )}\equiv
{
\left|\langle \exp\left(\mathrm{i}\Delta\phi/2\right)  \rangle\right| - \left|\langle \exp\left(\mathrm{i}\varsigma\phi/2\right) \rangle\right|},
\label{eq:5}
\end{eqnarray}
{ranging between $[-1,+1]$.}
With the aid of the concept of {wavefront} correlation degree introduced above, we can theoretically and numerically compute the correlation properties of various complex correlated light fields, particularly for a pair of conjugate {pseudo-thermal} light fields.

For the laser beam, we have $p^{(1)}=p^{( \frac{1}{2} )}=0$. For random light beams, it is assumed that the optical intensity follows a negative exponential distribution and that $A_1=A_2$, so the degree of intensity correlation always has $g^{(2)}(I_1,I_2)=2$. 
In Table~\ref{tab1}, we calculate correlation degrees for various random light sources, where $\mathrm{U}$ represents the uniform distribution. For two independent random light sources, $p^{(1)}=p^{( \frac{1}{2} )}=0$. Crucially, the sign serves as a quantitative classifier for the nature of {wavefront} correlation between pairwise correlated random beams.
\begin{table}[htbp]
\caption{Numerical calculation of various correlation degrees of random light sources. Except for $|g^{(1)}|$, the physical quantities listed in this table are usually difficult to measure directly in experiments.}
  \label{tab1}
  \centering
\begin{tabular}{ccccc}
\hline
Case & $|g^{(1)}|$ & $|h^{(1)}|$ & $p^{(1)}$ & $p^{(\frac{1}{2})}$ \\
\hline
$\phi_j\sim \mathrm{U}(0,2\pi)$ & $0$ & $0$ & $0$ & $0$ \\
$\phi_j\sim \mathrm{U}(0,2\pi)$\,$\Delta\phi=\mathrm{const.}$ & $1$ & $0$ & $1$ & $1$ \\
$\phi_j\sim \mathrm{U}(0,2\pi)$\,\,$\varsigma\phi=\mathrm{const.}$ & $0$ & $1$ & $-1$ & $-1$ \\
$\phi_j\sim \mathrm{U}(0,\pi)$\,\,\,$\Delta\phi=\mathrm{const.}$ & $1$ & $0$ & $1$ & $0.36$ \\
$\phi_j\sim \mathrm{U}(0,\pi)$\,\,\,\,$\varsigma\phi=\mathrm{const.}$ & $0$ & $1$ & $-1$ & $-0.36$ \\
\hline
\end{tabular}
\end{table}
Compared to $p^{(1)}$, the absolute value of $p^{(\frac{1}{2})}$ can also accurately describe the strength of phase fluctuations, and generally for $\phi_j\sim \mathrm{U}(0,\varphi)$, we obtain
\begin{eqnarray}
|p^{(\frac{1}{2})}|=1-\mathrm{sinc}(\varphi/2), \varphi\in[0,2\pi),
\label{eq:6a}
\end{eqnarray}
which indicates that $|p^{(\frac{1}{2})}|$ can reach its maximum value of 1 only if the phase fluctuates across the full range, so $p^{(\frac{1}{2})}$ is considered to be more informative.
 {In practice, the value of $\varphi$ can be flexibly controlled using an SLM \cite{Maga2016}.}
{
For pseudo-thermal light with a relatively long equivalent coherence time, the measurement of wavefront correlation degree is in principle feasible. For instance, a series of advanced wavefront sensing techniques \cite{Stoklasa2014, Soldevila2018, WuY2019, GaoY2026} can be employed to acquire the spatial wavefront information \(\phi_1\) and \(\phi_2\) in real time, which can then be substituted into Eqs. (\ref{eq:4}) and (\ref{eq:5}) for digital computation.}

More generally, when $\Delta\phi$ ($\varsigma\phi$) is not a constant but follows a certain statistical distribution [$\mathrm{P}_g(\Delta\phi)$ and $\mathrm{P}_h(\varsigma\phi)$], the degree of {wavefront} correlation can be further expressed as
\begin{equation}
p^{(1)}=\left| \int g^{(1)}\mathrm{P}_g(\Delta\phi) \mathrm{d}\Delta\phi \right| - \left| \int h^{(1)}\mathrm{P}_h(\varsigma\phi) \mathrm{d}\varsigma\phi \right|,
\label{eq:7}
\end{equation}
{where \(P_g(\Delta\phi)\) and \(P_h(\varsigma\phi)\) represent the probability distributions of the wavefront-difference \(\Delta\phi\) and wavefront-sum \(\varsigma\phi\) between a pair of pseudo-thermal light fields.}
The general expression of $p^{(\frac{1}{2})}$ is similar and not given here.

Furthermore, the proposed {wavefront} correlation degree can also be extended to various complicated pseudo-thermal light sources, such as the mixture of holographic thermal light {(i.e., a pair of conjugate fields $E_a$ and $E_a^*$)} and conventional thermal light {($E_b$)}.
{As shown in Fig.~\ref{fig0}(a)}, we consider the model of pairwise correlated sources between $E_1=E_a+kE_b$ and $E_2=E^*_a+kE_b$, where the statistical properties of $E_a$ and $E_b$ follow the complex Gaussian distribution and are statistically independent with identical average light intensity, and the real number $k$ is a scaling factor.
It should be noted that, in any given mixture of $E_j$, the two constituent elements are indistinguishable, as they are assumed by default to share the same frequency and polarization.
{Also, since both $E_a$ and $E_b$ follow the complex Gaussian distribution, both $E_1$ and $E_2$, obtained by linear superposition of $E_a$ and $E_b$, must obey the complex Gaussian distribution as well.}
Figure~\ref{fig0}(b) plots the curves of $p^{(\frac{1}{2})}$ and $p^{(1)}$ as functions of $k$. 
When $k=1$, the two types of fluctuations are balanced, resulting in $p^{(\frac{1}{2})}=p^{(1)}=0$. For $k > 1$, conventional thermal light prevails, consequently $p^{(\frac{1}{2})}$,$p^{(1)}>$0, with the converse being true for $k<1$.
{Therefore, the sign of $p^{(1)}$ or $p^{(\frac{1}{2})}$ can be used to quantitatively describe the wavefront correlation in pseudo-thermal light, i.e., whether it tends toward wavefront‑sum or wavefront‑difference correlation.
}

\subsection{Generalized Siegert relation for Gaussian {pseudo-thermal light}}

In Fig.~\ref{fig0}(c), numerical results also show that the original Siegert relation in Eq.~(\ref{eq:2c}) holds only when $k$ is sufficiently large; when $k$ approaches 0, it seems that the relation should be modified to $1+|h^{(1)}|^2$. For a genercal case of any value of $k$, Eq.~(\ref{eq:2c}) should be rewritten as a generalized form of $g^{(2)}=1+\left|g^{(1)}\right|^2+\left|h^{(1)}\right|^2$.
{
These results indicate that the classical Siegert relation does not apply to a pair of conjugate pseudo-thermal light sources and some complex mixtures, even though these optical fields obey complex Gaussian distributions.
These results also imply a perfect correction for it.
}

Now, based on complex Gaussian variables, we present a proof of the generalized Siegert relation here.
We consider two complex Gaussian monochromatic optical fields, defined as 
\begin{equation}
E_1 = e_1 + \Delta E_1,\quad E_2 = e_2 + \Delta E_2,
\label{eq1}
\end{equation}
where $e_1$ and $e_2$ are two constant values; $\Delta E_1$, $\Delta E_2$ are zero-mean fluctuation components, satisfying $\langle \Delta E_1 \rangle = \langle \Delta E_2 \rangle = 0$. 
Equation~(\ref{eq1}) indicates that this light field may follow a complex Gaussian distribution {probably} with a non-zero mean. Actual examples are those cases of the mixture of coherent light and thermal light \cite{OuZY1989, ChenH2011, LiuJ2014, ZHaoJL2026, DengX2026}. 

For brevity, the second-order moments of the fluctuation components are defined as 
\begin{equation}
G_{ij}\equiv\langle \Delta E_i \Delta E_j^* \rangle,\quad  H_{ij}\equiv\langle \Delta E_i \Delta E_j \rangle.
\label{eq2}
\end{equation}
Although each complex random variable $\Delta E_j$ individually follows a circular Gaussian distribution, the \textbf{joint} distribution of $\Delta E_1$ and $\Delta E_2$ does not necessarily satisfy the assumption of \textbf{circular symmetry}\cite{Gallager2008}.
Namely, for non-circularly symmetric jointly-Gaussian distributions, one has {$H_{11}$,$H_{22}=0$} but ${H_{12}\neq 0}$ (in contrast, $H_{12} = 0$ for circular symmetric ones), which is fundamentally different from all the assumptions made in classical {pseudo-thermal light sources} of the past. 
According to the Gaussian moment theorem\cite{Reed1962} or Wick's theorem, we have\cite{Lemieux1999}
\begin{equation}
\langle\Delta E_1\Delta E_2\Delta E^*_1\Delta E^*_2\rangle=G_{11}G_{22}+G_{12}G_{21}+H_{12}H_{21}.
\label{eq2_a}
\end{equation}
The actual manifestations of non-circular symmetry include conjugate light beams produced in nonlinear optical systems \cite{Yariv2003} and holographic thermal light\cite{YeZ2025, HouW2025, YeZ2026} in linear systems.
As noted before, this paper does not take into account the time-frequency terms of the light field and assumes that $\tau=0$ by default. Furthermore, the discussed complex conjugate light field here should only be applied to the spatial phase of the light field and should be independent of time inversion.

\begin{figure*}[htbp]%
\centering
\includegraphics[width=.95\linewidth]{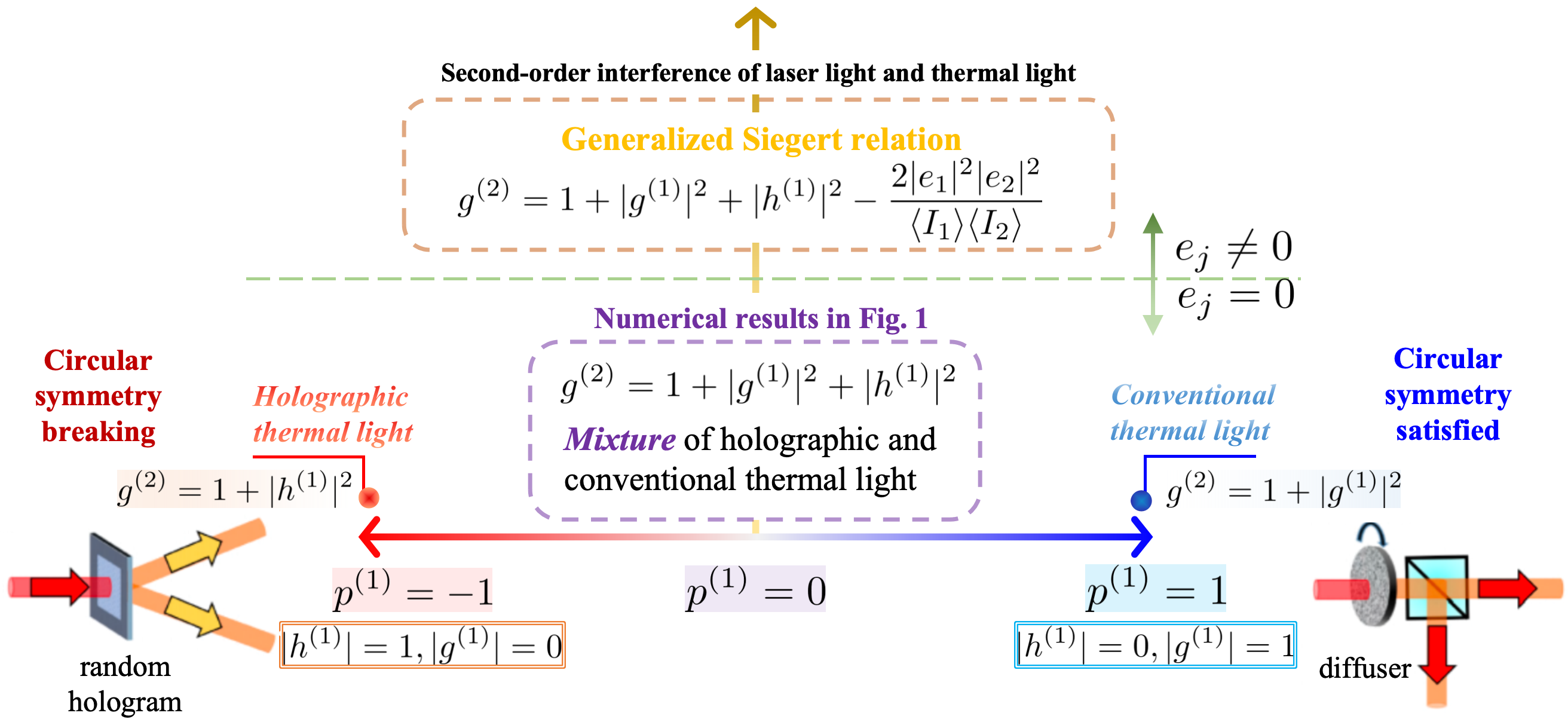}
\caption{Schematic diagram of the generalized Siegert relation. 
Based on whether $e_j$ is 0 and whether the non-zero jointly-Gaussian distribution satisfies circular symmetry, the generalized Siegert relation can be simplified into some common forms in Fig.~\ref{fig0}(c).
}\label{figa}
\end{figure*}

By substituting Eq.~(\ref{eq1}) into Eqs.~(\ref{eq:2a})(\ref{eq:3}), we have 
\begin{equation}
\begin{aligned}
|g^{(1)}|^2 \langle I_1 \rangle \langle I_2 \rangle = |e_1|^2|e_2|^2 + |G_{12}|^2 + 2\mathrm{Re}\left(G_{12} e_1 e_2^*\right),\\
|h^{(1)}|^2 \langle I_1 \rangle \langle I_2 \rangle = |e_1|^2|e_2|^2 + |H_{12}|^2 + 2\mathrm{Re}\left(H_{12} e_1^* e_2^*\right).
\end{aligned}
 \label{eq2_0}
\end{equation}

Based on the mean-fluctuation decomposition, the expectation of the optical field intensity can be derived as: $\langle I_1 \rangle = \langle E_1 E_1^* \rangle = |e_1|^2 + G_{11}$ and $\langle I_2 \rangle = \langle E_2 E_2^* \rangle = |e_2|^2 + G_{22}$. Then, the product of the two intensity expectations is 
\begin{equation}
\langle I_1 \rangle \langle I_2 \rangle = |e_1|^2|e_2|^2 + |e_1|^2 G_{22} + |e_2|^2 G_{11} + G_{11}G_{22}.
\label{eq2_1}
\end{equation}

Substituting Eq.~(\ref{eq1}) into the fourth-order moment $\langle I_1 I_2 \rangle = \langle E_1 E_1^* E_2 E_2^* \rangle$, according to the Gaussian moment theorem, we yield a general form
\begin{widetext}
\begin{equation}
\begin{aligned}
\langle I_1 I_2 \rangle & =\langle E_1 E_1^* E_2 E_2^* \rangle = \langle (e_1+\Delta E_1)(e_1^*+\Delta E_1^*)(e_2+\Delta E_2)(e_2^*+\Delta E_2^*) \rangle \\
&= |e_1|^2|e_2|^2 + |e_2|^2 G_{11} + |e_1|^2 G_{22} + G_{11} G_{22} + 2\mathrm{Re}\left(e_1^* e_2 G_{12} + e_1 e_2 H_{12}^*\right) + |G_{12}|^2 + |H_{12}|^2,\\
&= \langle I_1 \rangle \langle I_2 \rangle + 2\mathrm{Re}\left(e_1^* e_2 G_{12} + e_1 e_2 H_{12}^*\right) + |G_{12}|^2 + |H_{12}|^2.
\end{aligned}
 \label{eq5}
\end{equation}
\end{widetext}
where the function $\mathrm{Re}(\cdot)$ is to find the real part of the input complex value.
By substituting Eqs.~(\ref{eq2_a})-(\ref{eq5}) into Eq.~(\ref{eq:2b}), we can finally obtain the generalized Siegert relation as follows 
\begin{widetext}
\begin{equation}
\begin{aligned}
g^{(2)}=\frac{\langle I_1 I_2 \rangle}{\langle I_1 \rangle \langle I_2 \rangle} &= 1 + \frac{2\mathrm{Re}\left(e_1^* e_2 G_{12} + e_1 e_2 H_{12}^*\right) + |G_{12}|^2 + |H_{12}|^2}{\langle I_1 \rangle \langle I_2 \rangle},\\
&=\boldsymbol{1 + |g^{(1)}|^2 + |h^{(1)}|^2 - \frac{2|e_1|^2|e_2|^2}{\langle I_1 \rangle \langle I_2 \rangle}}.
\end{aligned}
 \label{eq6}
\end{equation}
\end{widetext}

\begin{figure*}[htbp]%
\centering
\includegraphics[width=0.96\linewidth]{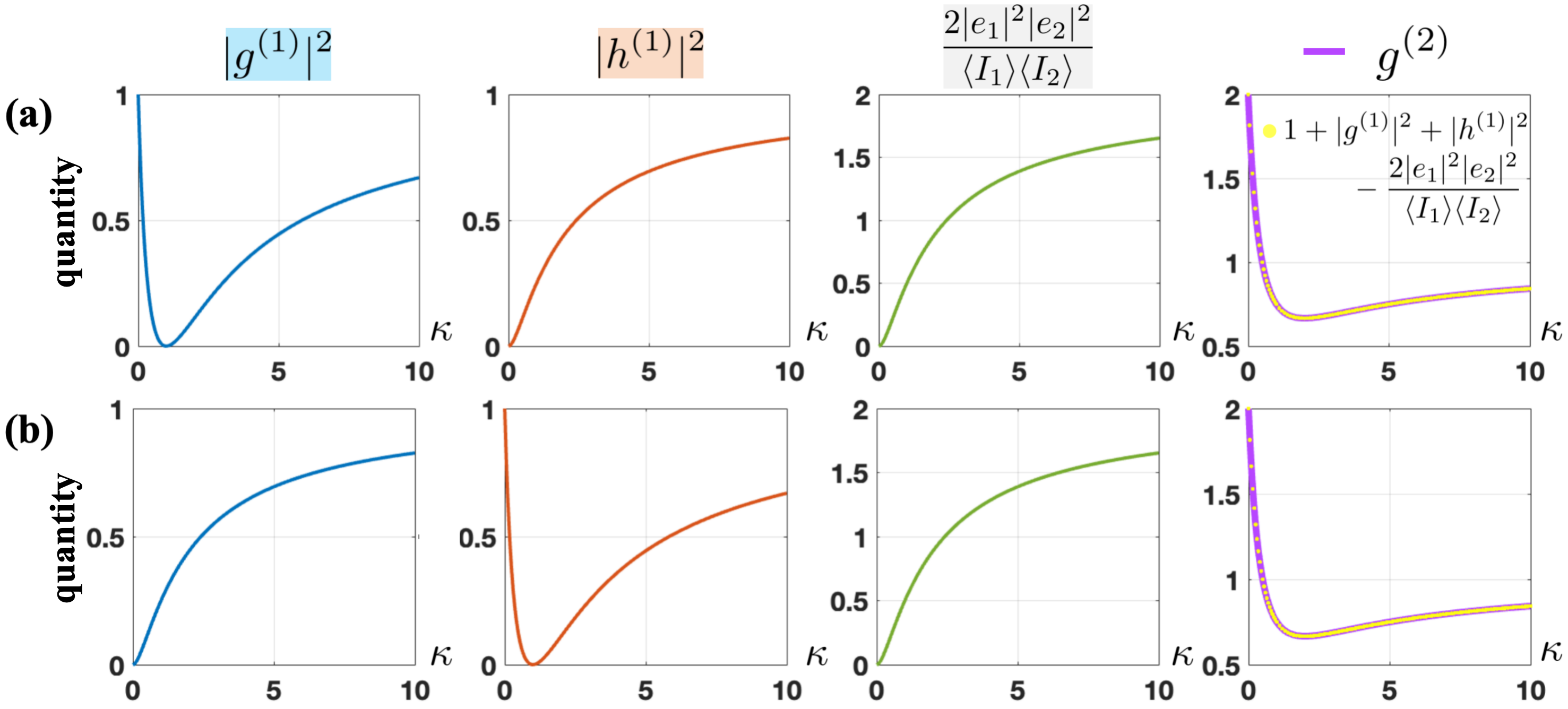} 
\caption{Numerical verification of the generalized Siegert relation between the interference fields of laser and thermal light in a beamsplitter.
(a) Second-order interference of laser light and conventional thermal light $\{\Delta E+a_0,\Delta E-a_0\}$; (b) Second-order interference of laser light and holographic thermal light $\{\Delta E+a_0,\Delta E^*-a_0\}$.
{The yellow dotted lines in the fourth column panel are the sum of the curves in the first three column panels plus 1.
}
}\label{figb}
\end{figure*}

Based on whether $e_j$ is 0 and whether the zero-mean jointly-Gaussian distribution satisfies circular symmetry, as shown in Eq.~(\ref{eq6}) and Fig.~\ref{figa}, the intensity correlation function $g^{(2)}$ as well as the Siegert relation can be simplified into some common cases below.
\begin{itemize}
\setlength{\itemsep}{1pt}
\setlength{\parskip}{0pt}
\setlength{\parsep}{0pt}
\item For the case of zero mean and circular symmetry [$|\langle E_j\rangle|=0$ and $H_{ij}=0$], we have the classic Siegert raltion of $\boldsymbol{g^{(2)}=1+|g^{(1)}|^2}$ given in Eq.~(\ref{eq:2c}) for conventional pseudo-thermal light.
\item For the case of zero mean and circular symmetry breaking [$|\langle E_j\rangle|=0$ and $H_{12}\neq0$], we have $\boldsymbol{g^{(2)}=1+|h^{(1)}|^2}$ corresponding to the blue dashed line in Fig.~\ref{fig0}(c) and the pratical case of holographic thermal light.
\item For the case of zero mean [$|\langle E_j\rangle|=0$], we have $\boldsymbol{g^{(2)}=1+|g^{(1)}|^2+|h^{(1)}|^2}$ that can well match with the numerical result in Fig.~\ref{fig0}(c) for the mixture of conventional thermal light and holographic thermal light.
\end{itemize}
Back to Fig.~\ref{fig0}(c), a particularly interesting and important case occurs when $k=1$, that is, when two types of thermal light are mixed in equal proportions. In this scenario, the degree of intensity correlation $g^{(2)}$ decreases to a minimum value of 1.5. 
By contrast, for the cases of $E_1=E_2=E_a+kE_b$ or $\{E_1=E_a+kE_b, E_2=E^*_a+kE^*_b \}$, representing coherent superposition of two independent sets of either conventional or holographic thermal light beams, it is always true for $g^{(2)}=2$ for any value of $k$.
It is also worth noting that the equal-proportion superposition of two independent, orthogonally polarized thermal lights also leads to a decrease in the intensity correlation degree to 1.5 \cite{Kellock2012, LiuXL2015, Kroh2020, HuZ2025}.

When considering Gaussian variables with non-zero means, however, all terms in Eq.~(\ref{eq6}) could be non-zero—for instance, the second-order interference of thermal light and laser light in a beam splitter.
For conventional thermal light, the interference fields can be expressed as $E_1=\Delta E + a_0$ and $E_2=\Delta E - a_0$ with $|a_0|^2=\kappa\langle|\Delta E|^2\rangle$, while we have $E_1=\Delta E + a_0$ and $E_2=\Delta E^* - a_0$ for holographic thermal light.
Figures~\ref{figb}(a)(b) display numerical results of various correlation degrees with the change of $\kappa$ in conventional and holographic thermal light, respectively.
It can be seen that the intensity correlation degree $g^{(2)}$ of the two types of pseudo-thermal light is consistent (third column), but the performance of $|g^{(1)}|^2$ (first column) and $|h^{(1)}|^2$ (second column) is exactly reversed.
Under the non-zero-mean condition, the fourth term in Eq.~(\ref{eq6}) is always negative, thus it is possible to observe the occurrence of anti-correlation phenomena, that is, the cross-correlation degree can even be less than 1. In this model of the second-order interference, the minimal value of $g_{\mathrm{min}}^{(2)}$ can reach $2/3$ when $\kappa=2$ as shown in Fig.~\ref{figb}, which is consistent with those experimental observations in Ref.~\cite{LiuJ2014}.

The results in Fig.~\ref{fig0}(c) verify the wide applicability of the generalized Siegert relation under zero mean and jointly non-circular symmetry, while the results in Fig.~\ref{figb} further confirm its applicability under non-zero mean. Overall, this generalized Siegert relation shall apply to any {pseudo-thermal light sources} that independently follow a complex Gaussian distribution, including their linear superposition {and the non-zero-mean condition}.
{Specifically, this model applies to any pair of Gaussian pseudo‑thermal light fields, including conventional Gaussian pseudo-thermal light, holographic thermal light, mixed sources of the two, and the second-order interference between either of them and laser light.}


\section{Experimental metric: correlation visibility}

Direct experimental observation of the proposed {wavefront} correlation degrees is challenging, unlike for $g^{(2)}$, due to its requirement for direct synchronous phase measurement of a pair of random optical fields. 
Moreover, the measurement of $h^{(1)}$ may require the use of nonlinear interferometers \cite{Chekhova2016}, {such as the SU(1,1) interferometer}.
In practical experiments, the most effective way to distinguish conventional thermal light from holographic thermal light is to observe the evolution of their spatial correlations \cite{YeZ2025, DingCX2025}. Holographic thermal light behaves similarly to entangled photon pairs, evolving from near-field position correlation to far-field momentum anti-correlation \cite{ChanKW2007, Tasca2009, R3, R4}. However, this observation method is only applicable to multimode random light. Also, it can be easily confused with other symmetric forms of conventional pseudo-thermal light \cite{YeZ2022, HRJ2025}, as such so-called momentum anti-correlation effects can also be readily simulated.

\begin{figure}[htbp]%
\centering
\includegraphics[width=\linewidth]{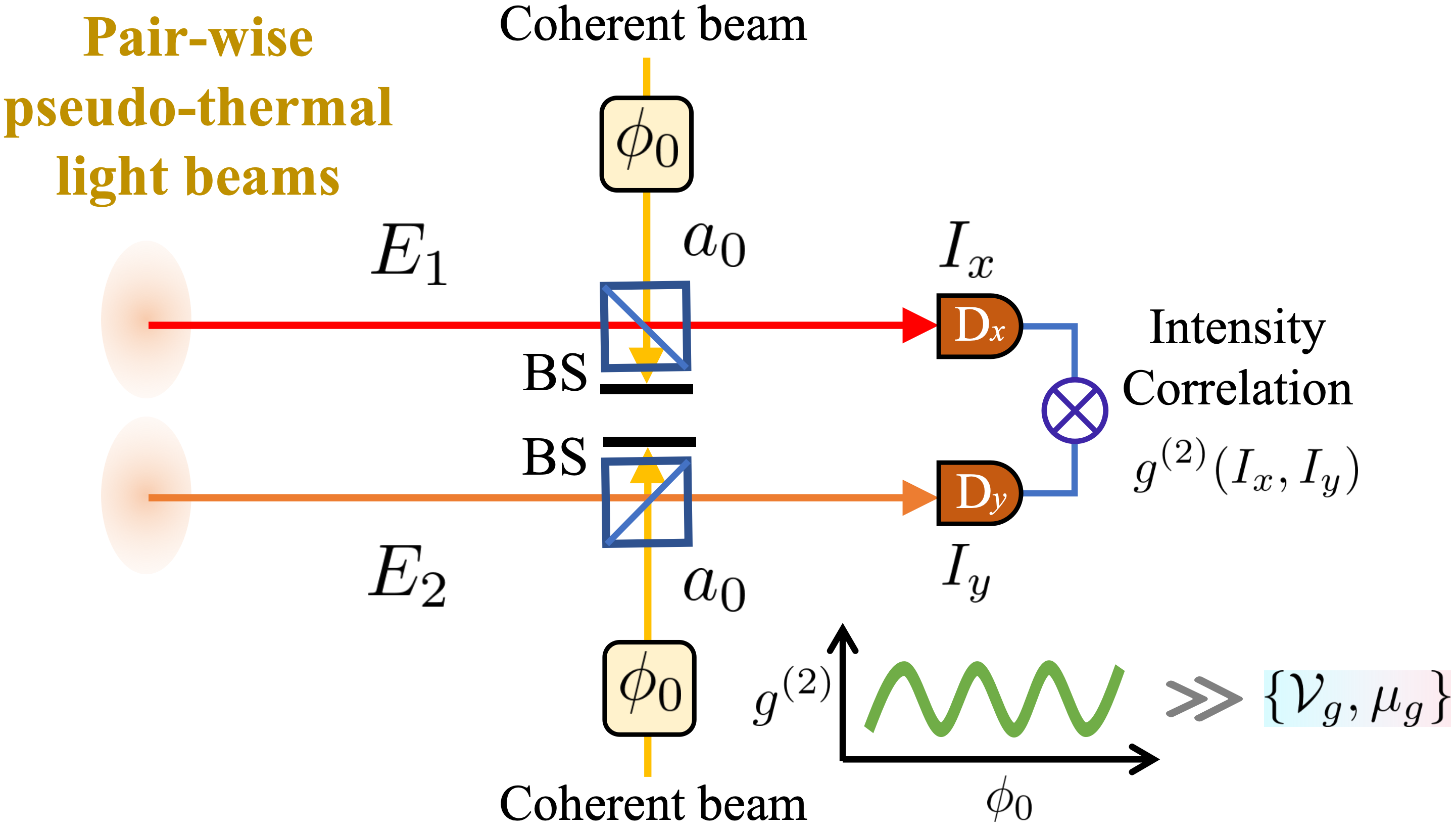}
\caption{Schematic diagram for measuring correlation visibility $\mathcal{V}_g$ and correlation background $\mu_g$ in an interferometer.}\label{fig1}
\end{figure}

To address this issue, another part of this work proposes an observable metric to further refine the generalized approach for calibrating various optical correlations between $E_1$ and $E_2$, where $E_j$ here is assumed to follow the complex Gaussian distribution with zero mean.
As shown in Fig.~\ref{fig1}, to allow the {wavefront} correlation to fully manifest, two identical coherent beams (local oscillators) $a_0=\sqrt{\mu_0}\exp(\mathrm{i}\phi_0)$ are introduced, with a relative adjustable phase shift $\phi_0$ and a real constant $\mu_0$ (i.e., light intensity), to obtain two coherently superposed fields, say $E_x=E_1+a_0$ and $E_y=E_2+a_0$. 
In fact, the situation discussed here is also a subset of Eq.~(\ref{eq1}) with $e_1=e_2=a_0$.

Notably, this process does not require quantitative phase measurement of the random light fields, but only the measurement of the two-point intensity correlation function $g^{(2)}(I_x,I_y)$ between these two interfering fields, where $I_x=I_1+\mu_0+2\sqrt{\mu_0I_1}\cos\left(\phi_1-\phi_0\right)$ and $I_y=I_2+\mu_0+2\sqrt{\mu_0I_2}\cos\left(\phi_2-\phi_0\right)$.
Based on the complex Gaussian variables, we have
$\langle I_x\rangle\langle I_y\rangle=(\mu_1+\mu_0)(\mu_2+\mu_0)$ and
$\langle I_xI_y\rangle=\langle I_1I_2\rangle+\mu_0^2+\mu_0(\mu_1+\mu_2)+2\mu_0\langle\sqrt{I_1I_2}\rangle\left[ \langle\cos(\varsigma\phi-2\phi_0)\rangle + \langle\cos(\Delta\phi)\rangle \right]$, where $\mu_j\equiv\langle I_j\rangle$. 
Hence, we have 
\begin{equation}
\begin{aligned}
g^{(2)}(I_x,I_y)=\frac{\left\langle I_xI_y\right\rangle}{ \left\langle I_x\right\rangle \left\langle I_y\right\rangle }=1+\frac{g^{(2)}(I_1,I_2)-1}{(\kappa_1+1)(\kappa_2+1)}\cdot\kappa_1\kappa_2 \\
+\frac{2\sqrt{\kappa_1\kappa_2}}{(\kappa_1+1)(\kappa_2+1)}\mathrm{Re}\left\{g^{(1)} + h^{(1)}\exp(-\mathrm{i}2\phi_0)\right\},
\end{aligned}
\label{eq:9}
\end{equation}
where $\kappa_j\equiv\mu_j/\mu_0$ represents the average intensity ratio of the Gaussian random light to the coherent light.
Equation~(\ref{eq:9}) reveals that $g^{(1)}$ and $h^{(1)}$ are coupled together. 
Note that $g^{(1)}$ or $h^{(1)}$ here is defined between $E_1$ and $E_2$, rahtar than between $E_x$ and $E_y$.
This coupling, albeit different from that defined for $p^{(1)}$, is capable of producing a noticeable and observable disparity.
For conventional thermal light (where $|g^{(1)}|=1$ and $|h^{(1)}|=0$), the value of $g^{(2)}(I_x,I_y)$ depends solely on $\kappa_j$.
However, when {wavefront}-sum correlation is present (i.e., $|h^{(1)}|$ is not identically 0), the value of $g^{(2)}(I_x,I_y;\phi_0)$ is also influenced by the initial phase $\phi_0$ of the coherent local oscillator and exhibits oscillations with two periods as $\phi_0$ varies, and in this case, the minimum value of  $g^{(2)}(I_x,I_y;\phi_0)$ is likely to be less than 1, demonstrating an anti-correlation effect.
As a result, the experimentally observed sinusoidal fringe in $g^{(2)}(I_x,I_y;\phi_0)$ can serve as direct evidence for the existence of {wavefront}-sum correlation.
Analogous to the definition of fringe visibility in Eq.~(\ref{eq:1}), we adopt the concept of \textit{correlation visibility}, 
\begin{equation}
\mathcal{V}_g=\frac{ g_{\mathrm{max}}^{(2)}-g_{\mathrm{min}}^{(2)} }{ g_{\mathrm{max}}^{(2)}+g_{\mathrm{min}}^{(2)} },
\label{eq:9a}
\end{equation} 
for further quantification for Eq.~(\ref{eq:9}):
\begin{equation}
\mathcal{V}_g=\frac{2\mathrm{Re}\left\{\int h^{(1)}\mathrm{P}_h \mathrm{d}\varsigma\phi\right\} }{ C_0 + \sqrt{\kappa_1\kappa_2} \left[g^{(2)}(I_1,I_2)-1\right]+ 2\mathrm{Re}\left\{\int g^{(1)}\mathrm{P}_g \mathrm{d}\Delta\phi\right\}},
\label{eq:10}
\end{equation}
where the constant $C_0\equiv(\kappa_1+1)(\kappa_2+1)/\sqrt{\kappa_1\kappa_2}$. It follows that $\mathcal{V}_g>0$ suggests a tendency for the pairwise correlated random lights to manifest the {wavefront}-sum correlation behavior. For a pair of conjugate random light beams or holographic thermal light ($|h^{(1)}|=1$ and $|g^{(1)}|=0$), the correlation visibility $\mathcal{V}_g$ attains its maximum value of $\sqrt{2}-1$ if and only if $\kappa_1=\kappa_2=1/\sqrt{2}$. In this case, the condition for maximizing $\mathcal{V}_g$ does not coincide with the condition for minimizing $g^{(2)}$.
In addition to visibility, the \textit{correlation background} $\mu_g$ of $g^{(2)}(I_x,I_y;\phi_0)$ in Eq.~(\ref{eq:9}) can be expressed as
\begin{equation}
\mu_g=1+\frac{g^{(2)}(I_1,I_2)-1}{(\kappa_1+1)(\kappa_2+1)}\kappa_1\kappa_2+\frac{2\sqrt{\kappa_1\kappa_2}\mathrm{Re}\left\{g^{(1)}\right\}}{(\kappa_1+1)(\kappa_2+1)}.
\label{eq:11}
\end{equation}
Now Eq.~(\ref{eq:9}) can be further fitted into a cosine function
\begin{eqnarray}
g^{(2)}(I_x,I_y;\phi_0)=\mu_g\cdot\left[1+\mathcal{V}_g\cos(-2\phi_0+\delta_0)\right],
\label{eq:12}
\end{eqnarray}
where $\delta_0$ is a constant initial phase shift related to the optical system itself.
Evidently, the correlation background $\mu_g$ is closely related to the degree of intensity correlation $g^{(2)}$, whereas the correlation visibility $\mathcal{V}_g$ is intimately linked to the {wavefront}-sum correlation. Together, both quantities $\{\mu_g,\mathcal{V}_g\}$ or $\{g^{(2)},\mathcal{V}_g\}$ can form a more complete, experimentally measurable two-dimensional classification scheme for various light sources as shown in Table~\ref{tab2}. Moreover, this 2D classification scheme can also be used to quantitatively characterize more complex pairwise random beams, similar to those presented in Fig.~\ref{fig0}. 
\begin{table}[htbp]
\caption{Numerical calculation of a two-dimensional classification scheme $\{\mu_g,\mathcal{V}_g\}$ or $\{g^{(2)},\mathcal{V}_g\}$ for various light sources. All physical quantities presented in this table are experimentally observable.}
  \label{tab2}
  \centering
\begin{tabular}{cccc}
\hline
Case & $\mathcal{V}_g$ & $\mu_g$ & $g^{(2)}$  \\
\hline
Laser beams & $=0$ & $=1$ & $=1$  \\
Two independent random beams  & $=0$ & $=1$ & $=1$ \\
Two identical random beams & $=0$ & $>1$ & $>1$  \\
A pair of conjugate random beams & $>0$ & $>1$ & $>1$  \\
\hline
\end{tabular}
\end{table}

\begin{figure*}[htbp]%
\centering
\includegraphics[width=0.75\linewidth]{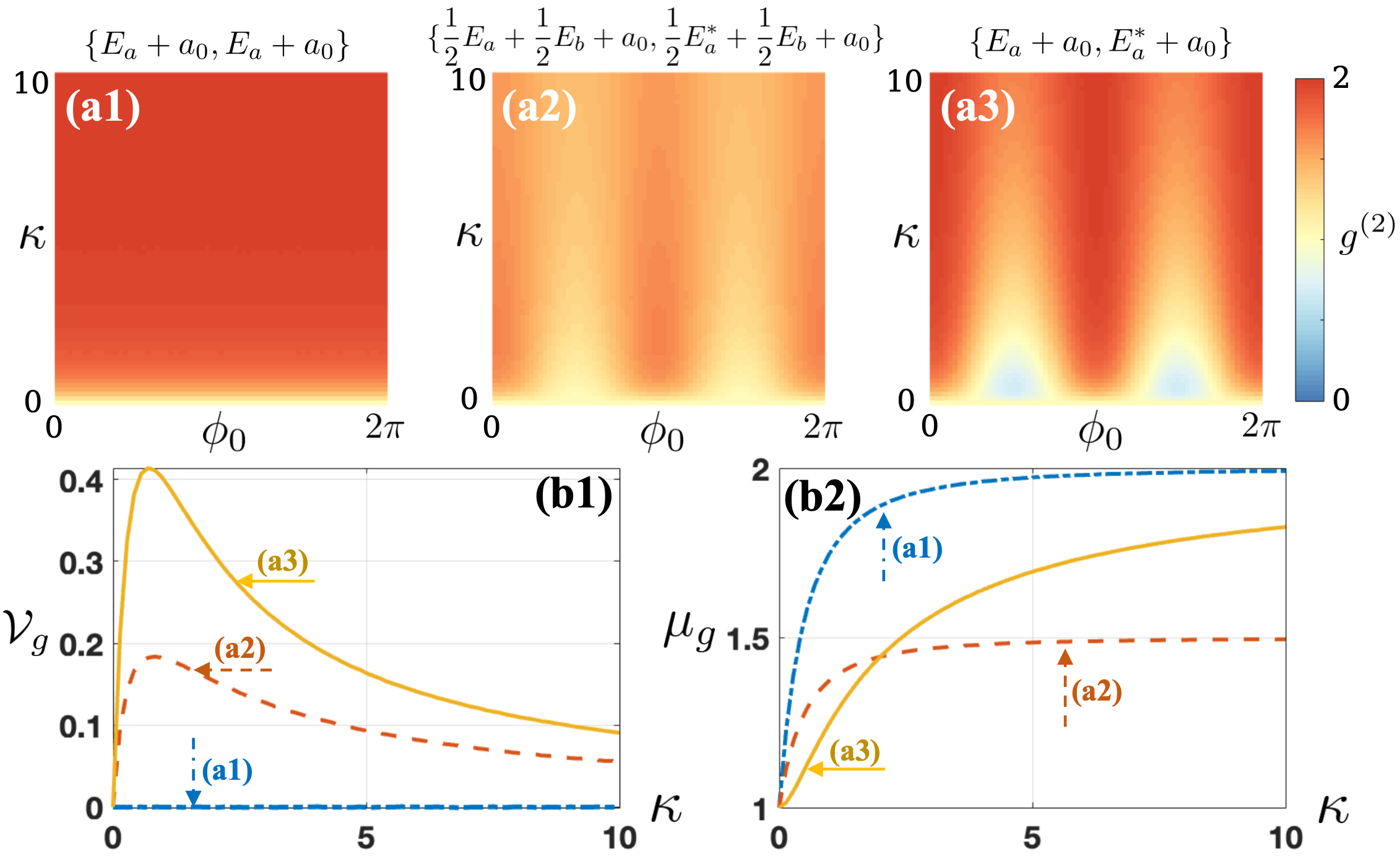}
\caption{Numerical calculation of intensity correlation degrees $g^{(2)}$, correlation visibility $\mathcal{V}_g$, and background $\mu_g$ for three types of pair-wise random light sources using the setup in Fig.~\ref{fig1}. From (a1) to (a3): 2D distributions of $g^{(2)}$ with the change of $\kappa$ ($\kappa_1=\kappa_2$) and $\phi_0$ corresponding to conventional thermal light ($E_x=E_a+a_0$, $E_y=E_a+a_0$), a mixture of conventional and holographic thermal light with an equal proportion ($E_x=0.5E^*_a+0.5E_b+a_0$, $E_y=0.5E_a+0.5E_b+a_0$), and holographic thermal light ($E_x=E^*_a+a_0$, $E_y=E_a+a_0$). (b1) and (b2) respectively depict the curves, corresponding to the three cases in (a1-a3), of the correlation visibility $\mathcal{V}_g$ and correlation background $\mu_g$ as $\kappa$ varies
 }\label{figc}
\end{figure*}

Based on the model shown in Fig.~\ref{fig1}, we numerically calculated the 2D distribution of the intensity correlation degrees $g^{(2)}$, as shown in Fig.~\ref{figc}, of conventional thermal light (a1), holographic thermal light (a3), and their equal-proportion mixture (a2) with respect to the changes of $\kappa$ and $\phi_0$ of the coherent beam (here we set $\kappa_1=\kappa_2$).
By comparing Fig.~\ref{figc}(a1) and Figs.~\ref{figc}(a2)(a3), when the jointly non-circular symmetry is not strictly satisfied, that is, when there are conjugate random light components, $g^{(2)}$ will oscillate sinusoidally with respect to $\phi_0$ as per Eq.~(\ref{eq:9}), ultimately resulting in a quantifiable visibility plotted in Fig.~\ref{figc}(b1).
On the other hand, the correlation background $\mu_g$ in Fig.~\ref{figc}(b2) cannot be regarded as a direct criterion for the jointly non-circular symmetry itself; This is because, as $\kappa$ increases, the correlation degrees of both traditional and holographic thermal light will trivially approach 2. Therefore, $\mu_g$ can be used, instead, as a criterion for the existence of intensity fluctuations, and can be employed to distinguish Gaussian random light from laser light.

\begin{figure}[htbp]%
\centering
\includegraphics[width=\linewidth]{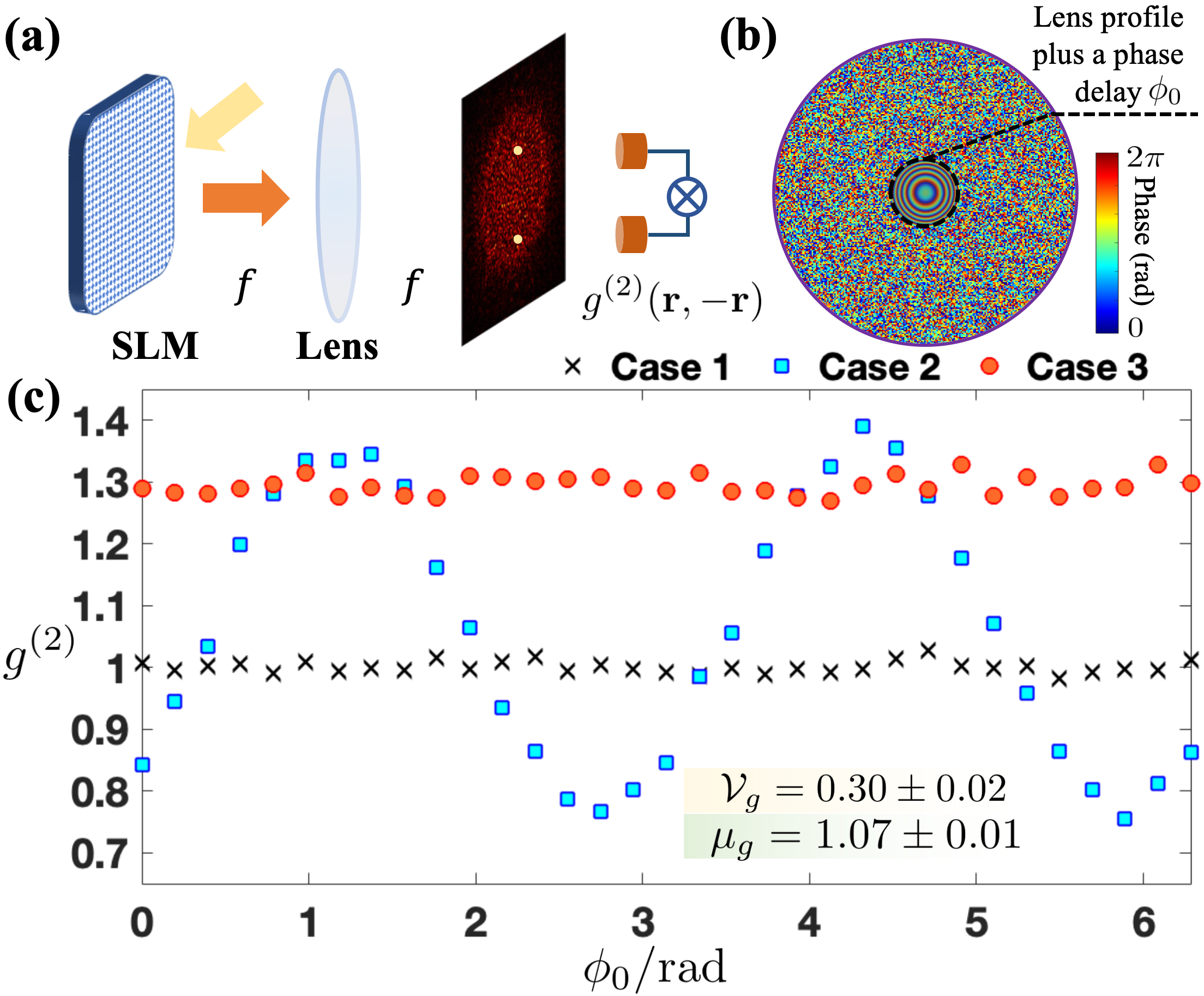}
\caption{Experimental observation of correlation visibility $\mathcal{V}_g$ and background $\mu_g$ for three types of random light sources. 
(a) Simplified experimental setup used to observe the intensity correlation degree $g^{(2)}(\mathbf{r},-\mathbf{r})$ between two centrosymmetric points, akin to Fig.~4 in Ref.~\cite{YeZ2022}; (b) Synthesized phase pattern used in the common-path interferometer; (c) Three plots of $g^{(2)}$ with the change of $\phi_0$. 
{$\phi_0$ takes 33 uniformly spaced values from 0 to $2\pi$.}
3000 realizations contribute to each value of $g^{(2)}$. The three cases correspond to conventional random beams, a conjugate pair of random beams, and an identical pair of random beams {(or centrosymmetric random beams)}, respectively.
 }\label{fig2}
\end{figure}

From an experimental perspective, $\{\mu_g,\mathcal{V}_g\}$ can be extracted by fitting the $\phi_0$-dependent oscillation of $g^{(2)}(I_x,I_y;\phi_0)$ based on Eq.~(\ref{eq:12}), or, more efficiently, via an analogous four-step phase-shifting approach.
For four definite phase shifts $\phi_0=0, \pi/4, \pi/2, 3\pi/4$, we then have
\begin{equation}
\begin{aligned}
\mu_g=\frac{1}{4}\left[g^{(2)}(0) + g^{(2)}(\frac{\pi}{4})  + g^{(2)}(\frac{\pi}{2})  + g^{(2)}(\frac{3\pi}{4}) \right], \\
\mathcal{V}_g=2\frac{\sqrt{ \left [ g^{(2)}(0) -g^{(2)}(\frac{\pi}{2}) \right]^2 + \left[ g^{(2)}(\frac{\pi}{4}) -g^{(2)}(\frac{3\pi}{4}) \right]^2 } }{ g^{(2)}(0) + g^{(2)}(\frac{\pi}{4})  + g^{(2)}(\frac{\pi}{2})  + g^{(2)}(\frac{3\pi}{4}) }.
\end{aligned}
\label{eq:13}
\end{equation}
It should be clarified that the four steps used here correspond to half of the four-step phase-shifting in digital holography \cite{Tahara2015}. This is because, in intensity correlation measurements, the phase $\phi_0$ of the coherent light is doubled as shown in Eq.~(\ref{eq:9}). In fact, a similar phenomenon can be observed in the angular spectrum transfer mechanism of holographic thermal light \cite{YeZ2025}, where the quadratic field of the incident light is imprinted onto the intensity correlation function.

For the observable correlation visibility $\mathcal{V}_g$ and correlation background $\mu_g$, we also provide a proof-of-principle experimental demonstration.
It is worth noting that the setup in Fig.~\ref{fig1} is essentially equivalent to introducing a coherent local oscillator into an HBT interferometer. Such an approach is quite common in the field of correlation optics \cite{Borghi2006,YeZ2023,TangX2025}.
Here, we adopt a common-path single-arm interferometer design \cite{YeZ2022,YeZ2023,Rosen2019}, as illustrated in Fig.~\ref{fig2}(a).
An expanded He-Ne laser beam (@632.8nm, the beam diameter is $\sim5$mm) is directed onto a commercial spatial light modulator (SLM, Holoeye-VIS-016, pixel pitch 8$\mu$m) loaded with varying holograms of a series of phase patterns displayed in Fig.~\ref{fig2}(b), in which a fixed diffractive lens with the focal length of 0.15m occupies the central circular region (the diameter is $\sim2$mm ) of the phase pattern. By contrast, the rest of the area is assigned a random phase distribution obeying a prescribed statistical profile, where each independent phase unit is composed of $5\times5$ pixels.

Three cases are investigated here: (1) a random phase distribution $\mathrm{U}(0,2\pi)$; (2) a binarized phase distribution that takes the value of 0 or $\pi$; (3) a centrosymmetric random phase distribution of $\mathrm{U}(0,2\pi)$, {$\phi(\mathbf{r})=\phi(-\mathbf{r})$}. An optical Fourier transform is carried out using a 0.3m focal-length Fourier lens, where a CMOS-based image sensor positioned at the back focal plane measures the intensity correlation between centrosymmetric points, i.e., $g^{(2)}(\mathbf{r},-\mathbf{r})$ (Refer to Ref.~\cite{YeZ2022} for more experimental details). 
{This pair of centrosymmetric points is taken vertically at a distance of approximately 0.9 mm from the center point.}

\begin{figure*}[htbp]%
\centering
\includegraphics[width=0.86\linewidth]{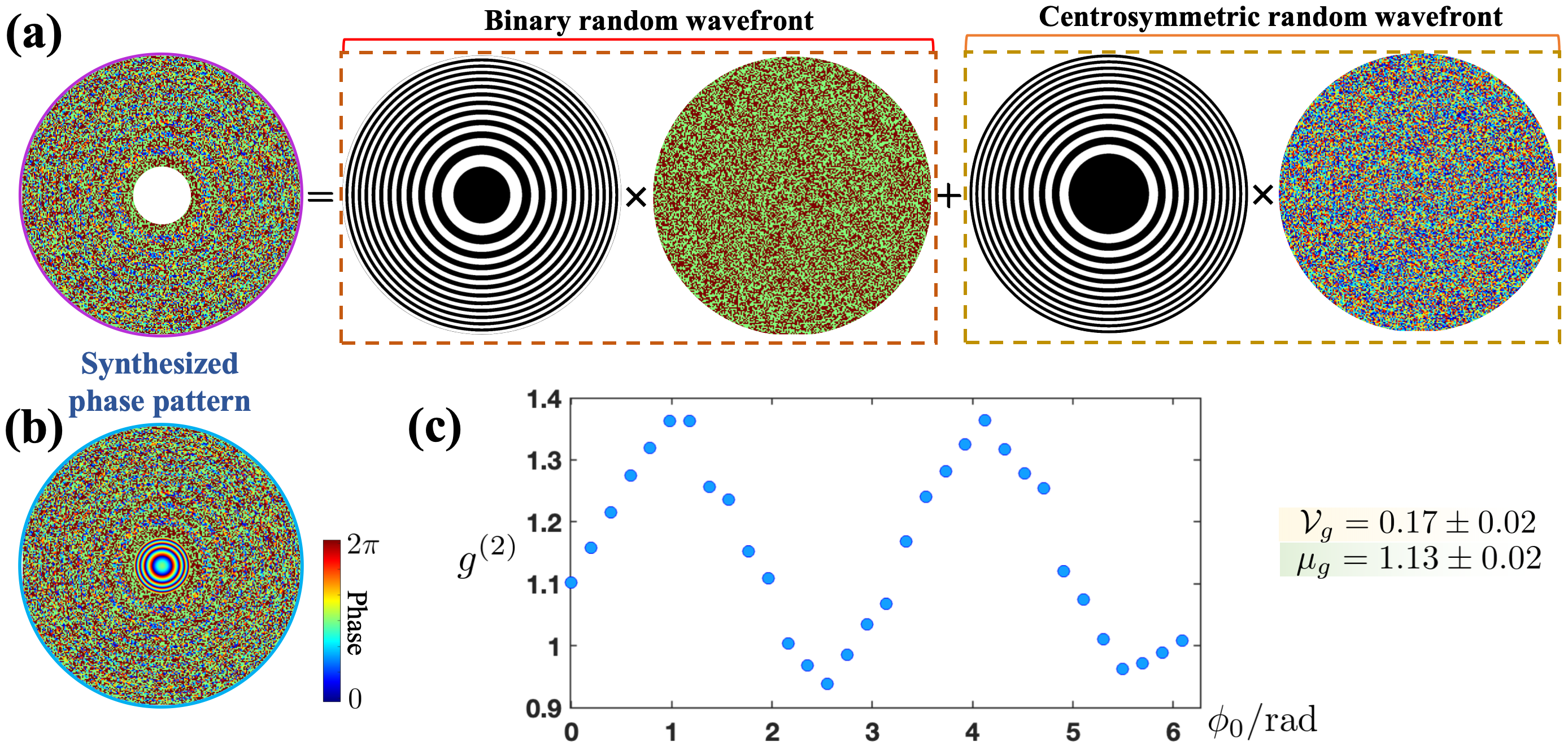}
\caption{
{
Experimental observation of intensity correlation degree $g^{(2)}$, correlation visibility $\mathcal{V}_g$, and background $\mu_g$ for the mixture of conventional and holographic thermal light (Case 4). The synthesized phase patterns in (a) and (b) are used to observe the intensity correlation degree $g^{(2)}(\mathbf{r},-\mathbf{r})$ and $\{\mathcal{V}_g,\mu_g\}$ of the mixed source, respectively. (c) shows the $\phi_0$-dependent curve of $g^{(2)}$ akin to Fig.~\ref{fig2}(c), confirming the existence of wavefront-sum correlation in the mixed source with $\mathcal{V}_g=0.17\pm0.02>0$.
}
 }\label{fig3}
\end{figure*}

In this setup, the speckle pattern in the Fourier plane arises from the interference of the reference spherical wave (generated by the central lens) with the random beams originating from the rest of the phase pattern. From the perspective of Fourier transform theory, the three designed phase distributions yield the following correlation characteristics at centrosymmetric positions: (i) conventional random beams (no intensity correlation between centrosymmetric positions), (ii) random beams with conjugate symmetry ({wavefront}-sum correlation), and (iii) random beams with even symmetry ({wavefront}-difference correlation).
The latter two cases will result in speckles with a centrosymmetric distribution in terms of intensity.
The SLM is imprinted with holograms corresponding to these synthesized phase patterns at a refreshing rate of 10 frames per second and synchronously triggers the image sensor to capture fluctuating speckle patterns.
A global phase shift $\phi_0$ is introduced to the central lens profile {ranging from 0 to $2\pi$}, and the intensity correlation degree $g^{(2)}$ between selected centrosymmetric points is measured and plotted in Fig.~\ref{fig2}(c) as a function of $\phi_0$. While offering robust stability, the common-path interferometer comes at the cost of flexible tuning of the intensity ratio $\kappa_j$ between the reference and the random light.
The experimental results confirm the findings in Table~\ref{tab2}. Case 1 is characterized by the absence of intensity correlation ($g^{(2)}=\mu_g=1$). In Case 2, the intensity correlation degree oscillates with $\phi$ [$g_{\mathrm{min}}^{(2)}\approx0.76$], yielding a visibility $\mathcal{V}_g${$=0.30\pm0.02$} and a background $\mu_g${$=1.07\pm0.01$} {by performing curve fitting of the experimental data according to Eq.~(\ref{eq:12}) and the nonlinear least-square approach}.
Case 3 shows no oscillation ($\mathcal{V}_g\approx0$) and has a background of {$1.30\pm0.02$}.
{Note that when the obtained experimental curve cannot be fitted by a cosine function, or the fitting error is too large (Case 1 and Case 3), the correlation visibility $\mathcal{V}_g$ is taken to be approximately 0.}
Based on the proposed common-path interferometer, these three types of random light sources can now be easily distinguished solely through joint correlation measurements and the 2D classification indicator $\{\mu_g,\mathcal{V}_g\}$. 
More significantly, the correlation visibility can serve as an experimentally observable criterion for the jointly non-circular symmetry in the zero-mean case.

 \begin{table}[htbp]
\caption{{Experimental observation of the intensity cross-correlation degree $g^{(2)}(\mathbf{r},-\mathbf{r})$ and  $\{\mu_g,\mathcal{V}_g\}$ for various random light sources.}}
  \label{tab3}
  \centering
\begin{tabular}{cccc}
\hline
Case & $\mathcal{V}_g$ & $\mu_g$ & $g^{(2)}$  \\
\hline
\makecell{Case 1: Two  \\ independent random beams}  & $0$ & $1$ & $1$ \\
\hline
\makecell{Case 2: A pair of \\ conjugate random beams } & ${0.30(2)}$ & $1.07(1)$ & $1.90$  \\
\hline
\makecell{Case 3: Two identical \\ random beams} & $0$ & $1.30(2)$ & $1.90$  \\
\hline
\makecell{Case 4: \\ The mixed source} & $0.17(2)$ & $1.13(2)$ & $1.46$  \\
\hline
\end{tabular}
\end{table}

{
Figure \ref{figc}(a2) numerically analyzes the correlation properties of a mixture of conventional and holographic thermal light. Here we experimentally observe such a mixed source (Case 4) in the setup of Fig.~\ref{fig2}(a), using the correlation visibility $\mathcal{V}_g$ and background $\mu_g$ that can be directly characterized in experiments. 
Similarly, we employ a common-path interferometer in which the phase pattern is designed using a spatial multiplexing strategy. As shown in Fig. \ref{fig3}(a), except for the central region (corresponding to the diffractive lens area, serving as the reference beam), the effective field of view is divided into multiple equal-area annular rings. Half of these rings are filled with a binarized random phase pattern to generate conjugate symmetric random beams, while the other half are filled with a centrosymmetric random phase pattern to generate centrosymmetric random beams.
We first measure the intensity correlation degree $g^{(2)}$, with the profile of the diffractive lens at the center removed. A series of phase patterns in Fig.~\ref{fig3}(a) is fabricated into holograms and loaded onto the SLM. 
Under coherent illumination, this is equivalent to the coherent superposition of a pair of conjugate symmetric random beams and a pair of identical random beams (approximately a 1:1 mixture). It should be noted that the two are statistically independent of each other.
The measured intensity correlation $g^{(2)}(\mathbf{r},-\mathbf{r})$ between centrosymmetric points reveals a correlation peak of only approximately 1.46, whereas the autocorrelation peak $g^{(2)}(0)$ reaches as high as 1.94. Meanwhile, when the phase patterns consist entirely of either binarized random or symmetric random phase patterns, the cross-correlation peak $g^{(2)}(\mathbf{r},-\mathbf{r})$ reaches 1.90.
The observed decrease in the $g^{(2)}(\mathbf{r},-\mathbf{r})$ from 1.90 to 1.46 is in good agreement with the theoretical and numerical results shown in Fig.~\ref{fig0}(c), i.e., the mixed source leads to a reduction in $g^{(2)}(\mathbf{r},-\mathbf{r})$ from 2 to 1.5.
The phase pattern shown in Fig.~\ref{fig3}(b) adds a central diffractive lens compared to (a). Next, we load these patterns and follow a procedure similar to that in Fig.~\ref{fig2}(c) to measure the $g^{(2)}$ profile shown in Fig.~\ref{fig3}(c) with the change of $\phi_0$ added to the diffractive lens.
Curve fitting is performed according to Eq.~(\ref{eq:12}), yielding $\mu_g=1.13\pm0.02$ and $\mathcal{V}_g=0.17\pm0.02$, indicating the presence of wavefront-sum correlation in this mixed source. Meanwhile, compared with Case 2 in Fig.~\ref{fig2}(c), the visibility is significantly reduced. This is because, according to Eq.~(\ref{eq:10}), the mixed source contains a conventional thermal light component, leading to a non-zero $g^{(1)}$ in the denominator. These results are in good agreement with the numerical calculations shown in Fig.~\ref{figc}(b1).
Table~\ref{tab3} summarizes the experimental observation results for the four cases.
 }

\section{Conclusion}
To conclude, this work presents a theoretical framework based on the {wavefront} correlation degrees $p^{(1)}$ and $p^{(\frac{1}{2})}$, complemented by a generalized Siegert relation applicable to complex Gaussian variables with non-zero mean and non-circular symmetry.
This work also proposes experimental observables of correlation visibility $\mathcal{V}_g$ and background $\mu_g$ in a linear system, where the {wavefront}-sum correlation in random beams can be quantitatively measured through the phase-dependent intensity correlation function. Collectively, they provide a more comprehensive framework for evaluating both first- and second-order coherence in pairwise-generated random light beams, particularly in systems where {wavefront}-sum correlation may arise or predominate. 
This set of physical quantities could in the future be extended to characterize the correlation properties of other degrees of freedom in Gaussian light \cite{Tervo2003,Martinez2024,HuZ2025,XiongJ2025}, such as polarization, orbital angular momentum, and frequency. Future research will also focus on the situation of non-Gaussian random light \cite{Lemieux1999,Bromberg2014,Starshynov2018,Els2024}.

\begin{acknowledgments}
We wish to acknowledge the support of the National Natural Science Foundation of China (12504411, 12274037), the Basic and Applied Basic Research Foundation of Guangdong Province (2026A1515011704), and the Fundamental Research Funds for the Central Universities.

The authors declare no conflict of interest. 
\end{acknowledgments}

\end{document}